\documentclass[%
reprint,
amsmath,amssymb,
 aps,
]{revtex4-2}

\usepackage{graphicx}% Include figure files
\usepackage{bm}% bold math
\usepackage{hyperref}% add hypertext capabilities
\usepackage{xcolor} % Include the package at the beginning of your document

\def\aap{A\&A}
\def\apj{ApJ}
\def\mnras{MNRAS}
\def\apjl{ApJL}

\def\prd{PRD}

\def\prl{PRL}

\def\nat{Nature}

\def\prc{PRC}

\begin{document}

\preprint{APS/123-QED}

\title{Impact of stratified rotation on the moment of inertia of neutron stars}

\author{Jonas P. Pereira$^{1,2,3}$, Tulio Ottoni$^{2,4}$, Jaziel G. Coelho$^{2,5}$, Jorge A. Rueda$^{6,7,8,9,10}$, and Rafael C. R. de Lima$^{11}$}
\email{e-mails to: jpereira@camk.edu.pl, jorge.rueda@icra.it}
\affiliation{$^{1}$ Departamento de Astronomia, Instituto de Astronomia, Geofísica e Ciências Atmosféricas (IAG), Universidade de São Paulo, São Paulo, 05508-090, Brazil}
\affiliation{$^{2}$ Núcleo de Astrofísica e Cosmologia (Cosmo-Ufes) \& Departamento de Física, Universidade Federal do Espírito Santo, Vitória, 29075-910, ES, Brazil}
\affiliation{$^{3}$Nicolaus Copernicus Astronomical Center, Polish Academy of Sciences, Warsaw, 00-716, Poland}
\affiliation{$^{4}$Universit\`a degli Studi di Ferrara, Via Saragat 1, I--44122 Ferrara, Italy}
\affiliation{$^{5}$Divis\~ao de Astrof\'isica, Instituto Nacional de Pesquisas Espaciais, S\~ao Jos\'e dos Campos, 12227-010, SP, Brazil}
\affiliation{$^{6}$ICRANet, Piazza della Repubblica 10, I-65122 Pescara, Italy}
\affiliation{$^{7}$ICRA, Dipartimento di Fisica, Sapienza Universit\`a  di Roma, Piazzale Aldo Moro 5, I-00185 Roma, Italy}
\affiliation{$^{8}$ICRANet-Ferrara, Dipartimento di Fisica e Scienze della Terra, Universit\`a degli Studi di Ferrara, Via Saragat 1, I--44122 Ferrara, Italy}
\affiliation{$^{9}$Dipartimento di Fisica e Scienze della Terra, Universit\`a degli Studi di Ferrara, Via Saragat 1, I--44122 Ferrara, Italy}
\affiliation{$^{10}$INAF, Istituto di Astrofisica e Planetologia Spaziali, Via Fosso del Cavaliere 100, 00133 Rome, Italy}
\affiliation{$^{11}$Universidade do Estado de Santa Catarina, Joinville, 89219-710,  Brazil}

\date{\today}

\begin{abstract}
\noindent Rigid (Uniform) rotation is usually assumed when investigating the properties of mature neutron stars (NSs). Although it simplifies their description, it is an assumption because we cannot observe the NS's innermost parts. Here, we analyze the structure of NSs in the simple case of ``almost rigidity,'' where the innermost and outermost parts rotate with different angular velocities. This is motivated by the possibility of NSs having superfluid interiors, phase transitions, and angular momentum transfer during accretion processes. We show that, in general relativity, the relative difference in angular velocity between different parts of an NS induces a change in the moment of inertia compared to that of rigid rotation. The relative change depends {\it nonlinearly} on where the angular velocity jump occurs inside the NS. For the same observed angular velocity in both configurations, if the jump location is close to the star's surface—which is possible in central compact objects (CCOs) and accreting stars—the relative change in the moment of inertia is close to that of the angular velocity (which is expected due to total angular momentum aspects. If the jump occurs deep within the NS, for instance, due to phase transitions or superfluidity, smaller relative changes in the moment of inertia are observed; we found that if it is at a radial distance smaller than approximately $40\%$ of the star's radius, the relative changes are negligible. Additionally, we outline the relevance of systematic uncertainties that nonrigidity could have on some NS observables, such as radius, ellipticity, and the rotational energy budget of pulsars, which could explain the X-ray luminosity of some sources. Finally, we also show that non-rigidity weakens the universal $I$-Love-$Q$ relations.
%\ubegin{description}
%\end{description}
\end{abstract}

\maketitle

\section{\label{sec:level1}Introduction}

A reasonable picture of the neutron star (NS) structure has emerged after more than half a century since the discovery of pulsars \cite{Haensel:2007yy}. First, there is an atmosphere of ionized matter, probably hydrogen or helium, with a composition that depends on the environment and can be affected by strong magnetic fields. Under the atmosphere, we have the ocean and then the crust, a lattice of increasingly heavy elements and neutron richness until the point of neutron drip density, which is energetically favorable to have free neutrons instead of being bound to nuclei. When the density increases, the lattice structure disappears, giving way to a liquid structure of primarily protons, neutrons, and electrons (outer core). Matter here is expected to be a superfluid. Going deeper into the inner core of an NS, several extra possibilities emerge, such as meson condensates \cite{mesoncondensate}, hyperons, and deconfined quarks \cite{Ivanenko:1965dg}. The core of an NS is where most of its mass is concentrated, and the energy density can be several times the nuclear saturation density. 

As all NS are spinning with frequencies ranging from sub-Hz \cite{2022NatAs...6..828C} to several hundreds of Hz \cite{fastpulsar, 2013FrPhy...8..679H}, it is crucial to describe appropriately the rotation and its effects on the star's observables. With such a rich and complicated internal structure, is treating a spinning NS as a rigid body reasonable? Numerical simulations show that newborn NS in mergers, supernova explosions, and main sequence stars all present differential rotation \cite{differentialrotation}. But even in an old and evolved NS, we can imagine different rotations between the crust and a weakly coupled superfluid core component due to vortices. Moreover, if there is a phase transition to quark matter in the neutron star core \cite{mixedphase}, simple angular momentum conservation considerations will demand a differential rotation between the crust and the now more compact core.

A clear example where rotation stratification is relevant is in the Sun. An abrupt change in angular velocity is possible due to the tachocline \citep{1992A&A...265..106S}, a very thin layer between the Sun's core and the convective envelope. In the case of NSs, the Ekman layer \citep{2006MNRAS.371.1311G} between the core and the crust may also lead to an abrupt change in the rotation of the phases it splits. Of course, stars also have dissipation mechanisms such as viscosity \cite{viscosity}, which leads to a uniform rotation inside some part of the star. However, there is still room for non-rigid rotation between two different phases of stars for those in the process of ``settling in'' (such as due to accretion) and those that are ``old'' (that could have superfluid parts). 

Regarding the viscosity, it has been estimated (see \citep{1994MNRAS.270..611L} and references therein for details) the characteristic timescales for an NS to reach uniform rotation. Assuming that superfluids are present in the outer core of an NS, the shear viscosity of the liquid core ($\eta$) would be dominated by electron-electron scattering ($\eta_{\rm{ee}}$), which is
\begin{equation}
    \eta_{\rm{ee}}=4.7\times 10^{19} T_8^{-2} \left( \frac{\rho}{\rho_{\rm {sat}}}\right)^2\rm{g\,cm^{-1}s^{-1}},
\end{equation}
where $\rho$ is the density, $T_8$ is the temperature in units of $10^8$ K, and $\rho_{\rm {sat}}=2.7\times 10^{14}$ g cm$^{-3}$ is the nuclear saturation density. In addition, $\eta\equiv \rho \nu$, where $\nu$ is the kinematic shear viscosity. Finally, the viscosity timescale can be estimated as $t_{\rm{vis}}\sim R^2/\nu$, where $R$ is the NS radius. From the above equation, one thus has
\begin{equation}
    t_{\rm vis}^{\rm {supfl}} = 0.18\,R_6^2\,T_8^2 \left( \frac{\rho}{\rho_{\rm sat}}\right)^{-1} \left(\frac{\eta}{\eta_{\rm{ee}}}\right)^{-1} \rm{yr},
\end{equation}
with $R_{6}=R/(10^6\,\text{cm})$. For instance, taking $T_8 = 1$, $R=10$ km and  $\rho= \rho_{\rm {sat}}$, it follows that $t_{\rm {vis}}^{\rm {supfl}}\approx 0.2$ yr. Thus, a superfluid phase in a star could be safely taken as having a uniform rotation. For the crust of an NS, analysis from \cite{2018MNRAS.481.4924Y} shows that $\eta_{\rm crust}\sim 10^{15}$ g cm$^{-1}$ s$^{-1}$ for $\rho_{\rm crust}=\rho_{\rm{sat}}/2$ and $T_8 = 1$. Thus, $t_{\rm {vis}}^{\rm {crust}}=R^2\rho_{\rm crust}/\eta_{\rm crust} \sim 10$ kyr, much longer than the superfluid region. Therefore, for old (i.e., aging more than the above timescale), non-accreting NSs, assuming rigid rotation of the crust is reasonable. When it comes to the timescale it takes for a superfluid phase and a non-superfluid phase to equilibrate their rotations, the answer is much more complex. Up to now, there is no precise mechanism for the core-crust angular momentum transfer. The occurrence of glitches suggests that the timescale for the rigid rotation of the whole star to be attained may be significant. Although the star will eventually reach an equilibrium angular momentum, this does not mean that angular velocities will be equal. This is because the superfluid part of the star (outer core and inner crust) is generally much larger in mass than the non-superfluid part (outer crust) \cite{Haensel:2007yy}.

This work aims to draw attention to the implications of rotation stratification of NSs, where we examine its impact on some relevant stellar observables. In general relativity (GR), the total angular moment $J$ appears as a constant of integration from the field equations \citep{1967ApJ...150.1005H} at the first order in the rotation parameter, associated with frame dragging. With the corresponding $J$, we find the moment of inertia $I$ by simply dividing it by the observed angular velocity $\Omega$ of the NS surface, measured, for example, with radio pulse timing, i.e., $I\equiv J/\Omega$. Note that this result is intrinsically general relativistic because $J$ comes directly from GR, and it generalizes the classical moment of inertia of stars with axial symmetry \citep{1967ApJ...150.1005H}. In the Newtonian framework, rotation generally influences the moment of inertia by altering the star's equilibrium shape, thereby affecting its mass distribution. However, for slow rotation (reasonable approximation for old NSs), the deviation from the moment of inertia in the static, spherically symmetric case remains small. On the other hand, when GR is also taken into account, boundary conditions for the angular velocity can have an imprint on $I$. That means that even in the perturbative case the background aspects of the star are enough in general. Thus, for NSs, the way internal parts rotate matters for the moment of inertia.

A relevant motivation for our study is that it will soon be possible to infer the moment of inertia directly from observations. For instance, further timing measurements of PSR J0737--3039A/B, the \textit{double pulsar}, will allow for further improvement in the constraints of the system post-Keplerian parameters, e.g., the periastron advance \cite{1988NCimB.101..127D, 2020ApJ...901..155G, 2021PhRvX..11d1050K}. Indeed, the inclusion of the pulsar A mass-energy loss owing to spindown will lead to a direct measurement of its moment of inertia with $11\%$ accuracy by 2030 \cite{2020MNRAS.497.3118H,2021PhRvX..11d1050K}. Another motivation is that the moment of inertia is important for several observables, such as the rotational energy, energy budget of stars, deformations, production of gravitational waves (GWs), the braking index and he physics of glitches \citep{Antonopoulou_2022}.

The structure of the paper is as follows: In Section \ref{slow rotation}, we give the basic details about slow rotation in general relativity, focusing on rigid rotation, differential rotation,  the correct boundary conditions when there is rotation stratification in NSs, the general relativistic version of the moment of inertia for stars and their rotational energy in this context. Section \ref{results} discusses some expected relative changes of the moment of inertia in the case of one-phase and hybrid stars when rotation stratification is present. Finally, in Sec. \ref{discussion}, we discuss the main points raised and show several astrophysical observables that could be affected by non-rigid rotation. We work with geometric units and metric signature $+2$ unless otherwise specified. 

%%%%%%%%%%%%%%%%%%%%%%%%%%%%%%%%%%%%%%%%%%%%%%%%%%%%%%
%%%%%%%%%%%%%%%%%%%%%%%%%%%%%%%%%%%%%%%%%%%%%%%%%%%%%%
\section{Slow rotation in GR, angular momentum and moment of inertia}
\label{slow rotation}
%%%%%%%%%%%%%%%%%%%%%%%%%%%%%%%%%%%%%%%%%%%%%%%%%%%%%%
%%%%%%%%%%%%%%%%%%%%%%%%%%%%%%%%%%%%%%%%%%%%%%%%%%%%%%

After Hartle and collaborators \cite{1967ApJ...150.1005H,1968ApJ...153..807H,1969ApJ...158..719H,1975ApJ...195..203H,1970ApJ...161..111H,1975ApJ...198..467H,1972ApJ...176..177H,1975ApJ...196..653H}, the problem of slowly rotating stars has been characterized and extensively explored. When it comes to the equations governing the motion of fluid elements and most metric components, they only appear in the second order of the rotation parameter $\Omega$, supposed to be small ($R\Omega \ll 1$, where $R$ is the stellar radius). However, the description of the angular rotation of the star's fluid, locally and for observers at infinity, is done in the first order of the rotation parameter. The structure of the star is described by the Tolman-Oppenheimer-Volkoff (TOV) equation, assuming spherical symmetry and a perfect-fluid description. 

In particular, the background spacetime is defined as 
\begin{equation}
    ds^2=-e^{\nu(r)}dt^2+e^{\lambda(r)}dr^2+r^2d\theta^2+r^2\sin^2\theta d\varphi^2.
\end{equation}
The unperturbed spacetime spherical symmetry allows the metric and variables decomposition into spherical harmonics ($Y^m_l$) or Legendre polynomials ($P_l$, $m=0$). 

\subsection{Rigid Rotation}
Up to the first order in the rotation parameter (for $l=1$), Hartle showed that the equation describing the rate of rotation of inertial frames, $\omega(r)$, is given by
\begin{equation}
    \frac{1}{r^4}\frac{d}{dr}\left(r^4j(r)\frac{d\omega}{dr} \right) -\frac{4}{r}\frac{dj}{dr}(\Omega - \omega )=0,\label{omega_equation}
\end{equation}
where $j(r)\equiv e^{-(\lambda+\nu)/2}$. In addition, $\Omega$ is the velocity of the fluid as seen by an observer at rest in the fluid. Equation \eqref{omega_equation} is a direct consequence of the $t\varphi$-component of the Einstein equations, which becomes relevant when deducing the boundary conditions in the stratified case. 

In rigid rotation, i.e., $\Omega$ constant, one could absorb it in the first term of the above equation and get \cite{1967ApJ...150.1005H}
\begin{equation}
    \frac{1}{r^4}\frac{d}{dr}\left(r^4j(r)\frac{d\bar{\omega}}{dr} \right) +\frac{4}{r}\frac{dj}{dr}\bar{\omega} =0,\label{omega_bar_equation}
\end{equation}
with $\bar{\omega} \equiv  \Omega - \omega$, the fluid's angular velocity relative to a freely falling observer, or, in other words, the fluid's angular velocity relative to the local inertial frame.

%%%%%%%%%%%%%%%%%%%%%%%%%%%%%%%%%%%%%%%%%%%%%%%%%%%%%
\subsection{Differential Rotation}
%%%%%%%%%%%%%%%%%%%%%%%%%%%%%%%%%%%%%%%%%%%%%%%%%%%%%

In the case of differential rotation, where $\Omega= \Omega(r)$ \citep{1970ApJ...161..111H}, the appropriate equation to be taken into account is \eqref{omega_equation}. In addition, to solve it, a prescription for $\Omega(r)$ should be given. Based on the sun's case \citep{1989ApJ...338..424P}, one would expect that $\Omega$ would not be the same throughout the star. However, if the star is old enough, it seems reasonable to assume rigid rotation given the action of angular momentum loss mechanisms. At the same time, that is exactly when superfluidity would play a role in the star's angular momentum. Thus, even for old stars, differential rotation should not be overlooked.  

Here, we do not work with differential rotation \textit{per se}, but we approach it by considering a toy model where some parts of the star rotate with different angular velocities. That should be seen as a phenomenological model trying to capture some of the rich physics in stars that would lead to a spatially varying rotation rate. This model should be applicable for some time intervals during the lives of compact stars. For instance, a young NS star born in a core-collapse supernova or an NS binary merger should experience some form of differential rotation \citep{Shapiro_2000}; it should also exist, at least transiently, in the case of fallback accretion (e.g., associated with central compact objects—CCOs—and recycled pulsars) \citep{2020MNRAS.499.3243S}, or even during an NS phase transition by angular momentum conservation.  

%%%%%%%%%%%%%%%%%%%%%%%%%%%%%%%%%%%%%%%%%%%%%%%%%%%%%
\subsection{Boundary Conditions} \label{boundary}
%%%%%%%%%%%%%%%%%%%%%%%%%%%%%%%%%%%%%%%%%%%%%%%%%%%%%

Given that, loosely speaking, the problem of perturbations of stars is a Sturm-Louiville problem, boundary conditions are paramount for their solutions. Hartle and collaborators have shown the necessary prescriptions for solving Eq. \eqref{omega_bar_equation}. Regularity at the origin imposes
\begin{equation}
    \bar{\omega} = {\rm const.},\;\;\; \left(\frac{d\bar{\omega}}{d r}\right)_{r=0}= 0.
\end{equation}
Outside the star, where $j=0$, one has a general solution
\begin{equation}
    \bar{\omega} = \Omega - 2\frac{J}{r^3},
\end{equation}
where $J$ is a constant identified with the total angular momentum of the star and in the absence of surface degrees of freedom, the function $\bar{\omega}$ and its derivative are continuous at the stellar surface. This implies that
\begin{equation}
    J=\frac{1}{6} R^4\left(\frac{d\bar{\omega}}{dr} \right)_{r=R},\;\;\; \Omega = \bar{\omega}(R)+ 2\frac{J}{R^3}.\label{J_and_Omega}
\end{equation}
Given one has freedom in choosing $\bar{\omega}$ at the center of the star, one will generally end up with a $\Omega$ different from a desirable one (coming from observations, for example). In this case, one should just re-scale $\bar{\omega}$ as follows
\begin{equation}
    \bar{\omega}_{\rm{new}}(r)= \left(\frac{\Omega_{\rm{new}}}{\Omega_{\rm{old}}}\right)\bar{\omega}_{\rm{old}}(r).\label{omega_new_old}
\end{equation}

When stratification is involved, further boundary conditions should be given. By stratification, we mean a sudden change in $\Omega$ at a given radial distance $R_{\star}$. Such a boundary condition could be obtained when promoting the relevant equations to distributions and collecting the Dirac delta terms. The appropriate equation to be promoted here is Eq. \eqref{omega_equation}. Assume that there is a sudden change of $\Omega$ at $r=R_{\star}$, meaning that at such a radial distance, the ``jump'' of $\Omega$ is different from zero, i.e., $[\Omega]^+_-\neq 0$. Here, ``$\pm$ '' represents a distance immediately above/below $r=R_{\star}$ ($r=R_{\star} \pm \epsilon$, with $\epsilon \rightarrow 0^+$). The question we want to answer is what $[\Omega]^+_-\neq 0$ implies for $[\omega]^+_-$ and $[\omega']^+_-$.

We stressed that it comes from the $t\varphi$-component of the Einstein equations, which should be the equation to be promoted to distributions. In this case, the energy-momentum tensor may have a surface term, contributing to a Dirac delta. Indeed, it appears when $[\omega']^+_-\neq 0$ \citep{2014PhRvD..90l3011P}. By writing (promotion of $\omega$ to a distribution)
\begin{equation}
    \omega = \omega^+ \Theta(r-R_{\star}) + \omega^- \Theta(R_{\star}-r),\label{omega_distribution}
\end{equation}
where $\Theta$ is the Heaviside function, and seeing Eq. \eqref{omega_equation} as a distributional equation with the right-hand side replaced by a term proportional to $[\omega']^+_-\delta (r-R_{\star})$, it immediately follows that this equation only starts making distributional sense if
\begin{equation}
    [\omega]^+_-=0.\label{jump_omega}
\end{equation}
Working now with the second derivatives of Eq.\eqref{omega_distribution}, based on Eq. \eqref{omega_equation} and the energy-momentum tensor at $r=R_{\star}$, it follows that
\begin{equation}
    [\omega']^+_-=0\label{jump_omega_derivative}.
\end{equation}
From the above equations, it is simple to see the appropriate jump conditions for $\bar{\omega}\equiv \Omega - \omega$:
\begin{equation}
    [\bar{\omega}]^+_-=[\Omega]^+_-,\;\;\; [\bar{\omega}']^+_-=0\label{jump_omega_bar}.
\end{equation}
We stress that it would have been incorrect to promote Eq. \eqref{omega_bar_equation} to a distribution and then find the appropriate boundary conditions to $\bar{\omega}$ because it is valid only when $\Omega$ is a given constant. Equations \eqref{jump_omega_bar} lead to the additional boundary conditions to be considered in the stratification case. One needs to solve Eq.\eqref{omega_bar_equation} for each layer with a given $\Omega$ and implement Eqs. \eqref{jump_omega_bar} between them.  

%%%%%%%%%%%%%%%%%%%%%%%%%%%%%%%%%%%%%%%%%%%%%%%%%%%%%
\subsection{Moment of Inertia in the stratified case}
%%%%%%%%%%%%%%%%%%%%%%%%%%%%%%%%%%%%%%%%%%%%%%%%%%%%%

In the rigid (uniform) perturbative rotation case without stratification, the moment of inertia is $I=J/\Omega$. This makes sense because the constant $J$, the star's total angular momentum, must be of first order in $\Omega$. It then ensues that $I$ should be seen as the general relativistic counterpart of the star's moment of inertia.

The fact that $I$ is a ratio of two first-order quantities does not mean it is free from the subtleties of $\bar{\omega}$, particularly its boundary conditions. To make matters more complicated, in the case of rotation stratification, it is in principle unclear which $\Omega$ (i.e., $\Omega^{\pm}$ in our case) to choose when writing $J$ as a first-order quantity. However, the case of uniform rotation and the observables we have at hand are good guides to consistently define $I$ for rotation stratification. To start with, for an isolated NS, the total angular momentum $J$ is a conserved quantity for both stratified and uniform rotation cases. In addition, the angular frequency at the surface of the star ($\Omega^+$) is an observable, easily obtained by timing analysis. Thus, the most natural definition for the moment of inertia in the stratified rotation case is also $I_{\rm{strat}} \equiv J/\Omega^+$.

This definition renders $I_{\rm{strat}}$ numerically different from $I_{\rm{rig}}$ in general because $J$ is sensitive to boundary conditions. However, the functional form of $J$ in the case of stratification is exactly the same as in the case of uniform rotation [given by Eq. \eqref{J_and_Omega}]. Indeed, from \citet{1970ApJ...161..111H} we learn that 
\begin{equation} 
J = -\frac{2}{3}\int^R_0 \frac{dj}{dr}r^3\bar{\omega}(r)dr= \frac{1}{6}R^4\left(\frac{d\bar{\omega}}{dr}\right)_{r=R}, \label{J_variational}
\end{equation} 
[see Eqs. (23), (28), (29), and (32) of \citet{1970ApJ...161..111H} for the case $l=1$, making use of $\int^0_{\pi}d\theta \sin^2\theta (dP_1(\theta)/d\theta) = -\frac{4}{3}$ with $P_1(\theta)$ the first-order Legendre polynomial and taking into account Eq. \eqref{omega_bar_equation}, $j(R)=1$ and $[\omega']^+_- = 0$]. From the above expression, one sees that different $J$s will arise in the case of stratified and rigid rotations because $\bar{\omega}_{\rm{strat}}'(R) \neq \bar{\omega}_{\rm{rig}}'(R)$. Note that (the constant) $J$ is intrinsically general relativistic, which easily allows us to define a relativistic moment of inertia using $\Omega^+$. In addition, $J$ is a ``global'' quantity, which depends on the internal aspects and dynamics of a star, in agreement with what is expected for its moment of inertia. Finally, $I = J/\Omega^+$ is physically relevant because $J$ (and also $\Omega^+$) can be directly inferred from observations \citep{1988NCimB.101..127D,2021PhRvX..11d1050K}.

%%%%%%%%%%%%%%%%%%%%%%%%%%%%%%%%%%%%%%%%%%%%%%%%%%%%%%%%%%
\subsection{Rotational energy}
%%%%%%%%%%%%%%%%%%%%%%%%%%%%%%%%%%%%%%%%%%%%%%%%%%%%

Another relevant issue for energy-balance considerations is the rotational energy of a star. Reference \cite{1970ApJ...161..111H} clarifies this point using the seminal work of Bardeen \cite{1970ApJ...162...71B}. In summary, Hartle has shown that the rotational energy (which is a positive quantity \cite{1970ApJ...161..111H}) of a star with uniform rotation (for which case $\omega_l=0$ for $l>1$) is given by
\begin{equation}
    E_{\rm{rot}}=-\frac{1}{3}\int^R_0dr r^3\frac{dj}{dr}\Omega(\Omega -\omega(r))\label{rot_energy},
\end{equation}
where it is now understood that $\Omega = \Omega^{+}\Theta (r-R_{\star})+ \Omega^{-}\Theta(R_{\star}-r)$, which is the generalization of the uniform rotation case for stratification. Since Eq. \eqref{rot_energy} comes from a variational principle, it should also hold in the case of stratification. However, its integrant is not continuous in the whole star, and thus this integral should be properly split. One has
\begin{equation}
     E_{\rm{rot}}=-\frac{1}{3}\left[\Omega^-\int^{R_{\star}}_0dr r^3\frac{dj}{dr}\bar{\omega}^- + \Omega^+\int_{R_{\star}}^R dr r^3\frac{dj}{dr}\bar{\omega}^+ \right]\label{rot_energy_parts},
\end{equation}
where we have used that $\bar{\omega}(r)=\Omega - \omega(r)$ for each phase. With Eq. \eqref{rot_energy_parts}, one can use Eq. \eqref{omega_bar_equation} and obtain
\begin{equation}
    E_{\rm{rot}}= \frac{1}{2}\left[J\Omega^+ -\frac{1}{6}R_{\star}^4j(R_{\star})\bar{\omega}'(R_{\star})[\Omega]^+_-\right].
\end{equation}

Therefore, from our definition of the moment of inertia,
\begin{equation}
    E_{\rm{rot}}= \frac{1}{2}\left[ I -\frac{1}{6}R_{\star}^4j(R_{\star})\frac{\bar{\omega}'(R_{\star})}{\Omega^+}\frac{[\Omega]^+_-}{\Omega^+}\right](\Omega^+)^2 \label{Erot_strat}.
\end{equation}
We stress that the second term of the above equation is generally much smaller than the first due to $R_{\star} < R$, $j(R_{\star}) < 1$, and ${|\,[\Omega]^+_-\,|}/{\Omega^+} < 1$. In general, it would have been misleading to read off the moment of inertia from Eq. \eqref{Erot_strat} because energy is not trivially extended from Newtonian dynamics to general relativity in general. Finally, we note that Eq. \eqref{Erot_strat} is only valid for the case of `almost rigidity," i.e., when $\Omega^{\pm}$ are constants. The case where $\Omega^{\pm} = \Omega^{\pm}(r)$ is much more complicated because  it involves $l > 1$ \citep{1970ApJ...161..111H}. We leave this to be studied in future work.

%%%%%%%%%%%%%%%%%%%%%%%%%%%%%%%%%%%%%%%%%%%%%%%%%%%%%
%%%%%%%%%%%%%%%%%%%%%%%%%%%%%%%%%%%%%%%%%%%%%%%%%%%%%
\section{Results}
\label{results}
%%%%%%%%%%%%%%%%%%%%%%%%%%%%%%%%%%%%%%%%%%%%%%%%%%%%%
%%%%%%%%%%%%%%%%%%%%%%%%%%%%%%%%%%%%%%%%%%%%%%%%%%%%%

Our main goal here is to compare the outcomes of $J$ and $I$ in the cases of rigid rotation and stratification. For simplicity, in the case of stratification, we assume that two regions of a star rotate with different $\Omega$s and that a surface splits them at $r=R_{\star}$. The case with more layers with different values of $\Omega$ can be trivially extended.

The motivations for different parts of the star rotating with different angular velocities stem from various stellar possibilities: (i) sharp phase transitions leading to a quark core and a hadronic phase (crust); (ii) superfluidity in the neutron star, which tends to preserve a certain angular velocity up to a critical lag; (iii) burying of the star's surface by supernova remnant material, which in general has different angular momentum than the star's; (iv) nuclear reactions and gravitational and electromagnetic wave emission in some regions of the star, which also take away angular momentum.

%%%%%%%%%%%%%%%%%%%%%%%%%%%%%%%%%%%%%%%%%%%%%%%%%%%%%
\subsection{Modeling}
%%%%%%%%%%%%%%%%%%%%%%%%%%%%%%%%%%%%%%%%%%%%%%%%%%%%%

We will work with two main models: (i) the SLy4 equation of state (EOS) \citep{2001A&A...380..151D} for one-phase stars and (ii) stars presenting sharp phase transitions with different quark-hadron energy density jumps ($\eta +1$)--hybrid stars. Many models could be used for the hybrid star EOS. Here we limit to some examples given in \citet{2023ApJ...950..185P}. Particular details can be found there. For the $M-R$ relations (or $M-R$ diagrams) of the representative EOSs that we will make use of here, see  Fig. \ref{fig0}. 

In case (i), we assume that $R_{\star}$ is chosen at will. It would allow one to build intuition about changes in $J$ and $I$ for different angular velocity stratification depths compared to the rigid rotation scenario. In case (ii), $R_{\star}$ will be identified with the phase transition radius and hence will be fixed for given quark and hadronic equations of state and stellar masses.  The idea of this work is not to make an exhaustive EOS analysis but to find the main trends to focus on in future works and to make back-of-the-envelope estimates of the relative changes relative to the rigid rotation case.

%%%%%%%%%%%%%%%%%%%%%%%%%%%%%%%%%%%%%%%%%%%%%%%%%%%%%
\subsection{One-phase stars with rotational stratification}
%%%%%%%%%%%%%%%%%%%%%%%%%%%%%%%%%%%%%%%%%%%%%%%%%%%%%

Let us focus first on stars without phase transitions described by the (realistic) SLy4 model. We will assume that the star's inner and outer parts rotate with different $\Omega$s. The inner part encompasses radii up to $R_{\star}$, which will be freely chosen. The outer part goes from $R_{\star}$ to the star's surface ($R$). Motivated by the sun's case, where relative changes in $\Omega$ for the core and the convective envelope could be around $10\%-20\%$ \citep{2009LRSP....6....1H} (associated with the presence of a very thin layer where the angular velocity changes rapidly--the tachocline \citep{1992A&A...265..106S}), we will assume this also to be the case of NSs roughly. Because we work with $\bar{\omega}$ in the numerical integrations, we will assume that  $[\Omega]^+_-/\bar{\omega}^- = \rm{given} \equiv {\cal C}$. From it, it trivially follows that  $[\Omega]^+_-/\Omega^+= {\cal C}(\bar{\omega}^-/\Omega^+)$. Here, $\Omega^{-}(\Omega^{+})$ is the angular velocity of the inner(outer) phase of the star. We aim to work with cases where $[\Omega]^+_-/\Omega^+\sim 10\%-20\%$. For a given ${\cal C}$, Eq. \eqref{omega_bar_equation} could be easily integrated out, and $\bar{\omega}$ could be known throughout the star. For a given $\Omega^+$, which directly comes from observations of NS surfaces, one could easily find the right ${\cal C}$ for any given $[\Omega]^+_-/\Omega^+$. As a rule of thumb, ${\cal C}$ is not much different from $[\Omega]^+_-/\Omega^+$. 

An important issue when comparing $I$ and $J$ with and without rigid rotation is fulfilling Eqs. \eqref{J_and_Omega} and \eqref{omega_new_old} if the $\Omega$ that we get at the star's surface is not the one we would like to link with observations. That is not the case, given that $\Omega$ at $R$ depends on the choice of $\bar{\omega}(0)$. However, this is not a problem, as we show now. From our definition of $I$ and Eqs. \eqref{J_and_Omega} and \eqref{omega_new_old}, it follows that $I=J_{\rm{new}}/\Omega_{\rm{new}}=J_{\rm{old}}/\Omega_{\rm{old}}$, meaning that it is independent of the choice of $\bar{\omega}$ at the center of the star. For $J$ in both the stratified and rigid rotation cases, one can use Eq. \eqref{J_variational} after solving Eq. \eqref{omega_bar_equation}. Note from Eq. \eqref{J_variational} that when $R_{\star}\rightarrow R$, it follows that $J_{\rm{strat}}\rightarrow J_{\rm{rig}}$ due to Eq. \eqref{jump_omega_derivative}.

\begin{figure}
\centering
\includegraphics[width=\columnwidth]{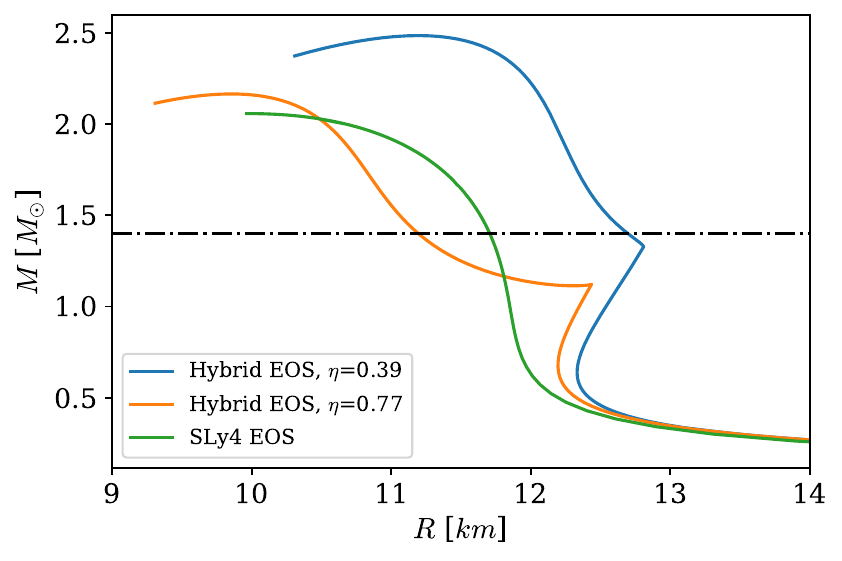}
\caption{$M$-$R$ relations for the SLy4 EOS and the hybrid models of Ref. \cite{2023ApJ...950..185P}, for selected values of the $1+\eta\equiv$ top of the quark phase over the bottom of the hadronic phase energy density ratio, $0.39$ and $0.77$. The dot-dashed horizontal line represents $M=1.4M_{\odot}$ and is highlighted to facilitate the identification of the canonical NS mass radius.}
\label{fig0}
\end{figure}

\begin{figure}
\centering
\includegraphics[width=\columnwidth]{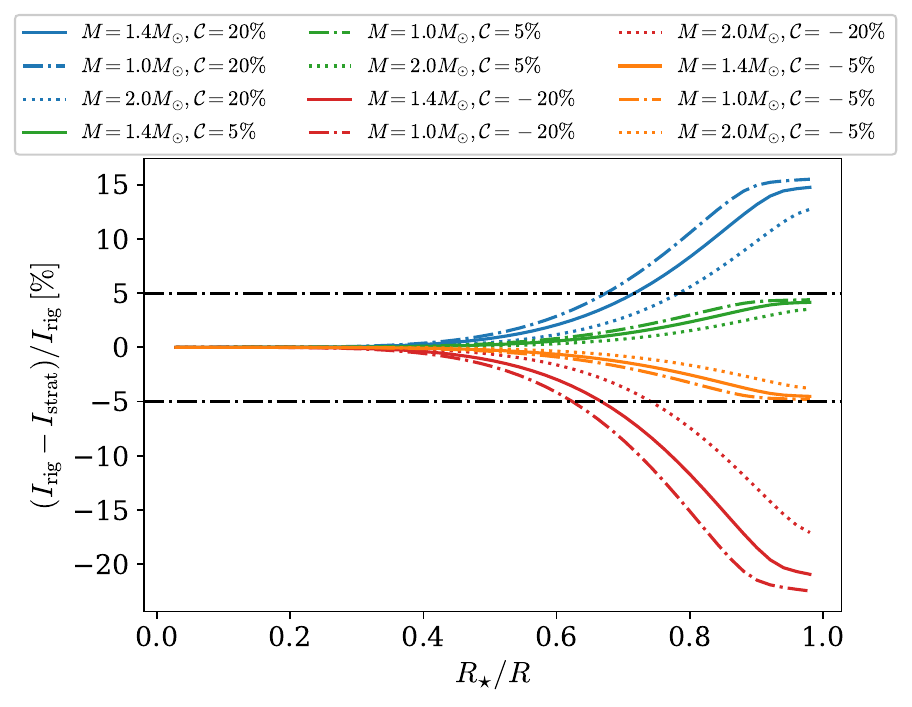}
\caption{$\Delta I/I_{\rm{rig}}$ for the SLy4 EOS assuming a relative angular velocity jump ${\cal C}$ at $R_{\star}$. Several ${\cal C}$s have been included, as well as NS masses. Each color represents a given ${\cal C}\sim [\Omega]^+_-/\Omega^+$, while each curve dashing relates to a given NS mass. Non-negligible relative differences start showing up for $R_{\star}/R\gtrsim 0.6$, increasing nonlinearly. If we can infer an NS moment of inertia with a $5\%$ uncertainty, that would also be the minimum level of uncertainty for the relative angular velocity difference between the two phases we could resolve.}
\label{fig1}
\end{figure}

We start by analyzing the relative differences of $I$ between stratification and rigid rotation for different depths in the NS (different $R_{\star}$). We use as a representative EOS the SLy4 EOS (see Fig. \ref{fig0} for its $M-R$ relation). 
Figure \ref{fig1} shows that the largest differences in $\Delta I/I_{\rm{rig}}\equiv (I_{\rm{rig}}-I_{\rm{strat}})/I_{\rm{rig}}$ happen close to the surface of the star ($\sim [\Omega]^+_-/\Omega^+$). This is expected since the total angular momenta in the cases of stratified and uniform rotation tend to the same value, meaning that $\Delta I/I_{\rm{rig}}\rightarrow [(1/\Omega^-)-(1/\Omega^+)]\Omega^-=[\Omega]^+_-/\Omega^+$. What is not intuitive is how it does so. It decreases nonlinearly when the angular velocity jumps deeper inside the star. Further, the smaller the mass, the larger the relative difference for a given $R_{\star}$. Roughly speaking, relative differences in $I$ grow much quicker and become relevant for $R_{\star}\gtrsim 0.6 R$. The sign of $\Delta I/I_{\rm{rig}}$ is always the same as $[\Omega]^+_-/\Omega^+$, but the relative changes in the moment of inertia are not exactly symmetric to the relative changes in $\Omega$. 

%%%%%%%%%%%%%%%%%%%%%%%%%%%%%%%%%%%%%%%%%%%%%%%%%%%%%
\subsection{Hybrid stars}
%%%%%%%%%%%%%%%%%%%%%%%%%%%%%%%%%%%%%%%%%%%%%%%%%%%%%

We consider EOSs with different (and controllable) energy density jumps for hybrid stars as in \cite{2020ApJ...895...28P, 2021ApJ...910..145P}. In other words, we use the SLy4 EOS \citep{2001A&A...380..151D} for the crust, smoothly connected to a polytropic EOS for the hadronic outer core, followed by a sharp phase transition with a given energy density jump, $\eta +1$, to an inner quark core modeled by a simple MIT bag-like model, with the sound speed equal to unity (as suggested by Bayesian inferences using GW and electromagnetic data \cite{2021PhRvC.103c5802X,2021ApJ...913...27L,2021PhRvD.104f3003L}). Further details about these EOSs can be found in Ref. \cite{2023ApJ...950..185P}. The mass-radius relations for the hybrid EOSs we use are shown in Fig. \ref{fig0} and they lead to radii around $11$--$13$ km for $1.4M_{\odot}$, in agreement with multimessenger results \cite{2019ApJ...887L..24M,2019ApJ...887L..21R,2021ApJ...918L..28M,2021ApJ...918L..27R}.

\begin{figure}
\centering
\includegraphics[width=\columnwidth]{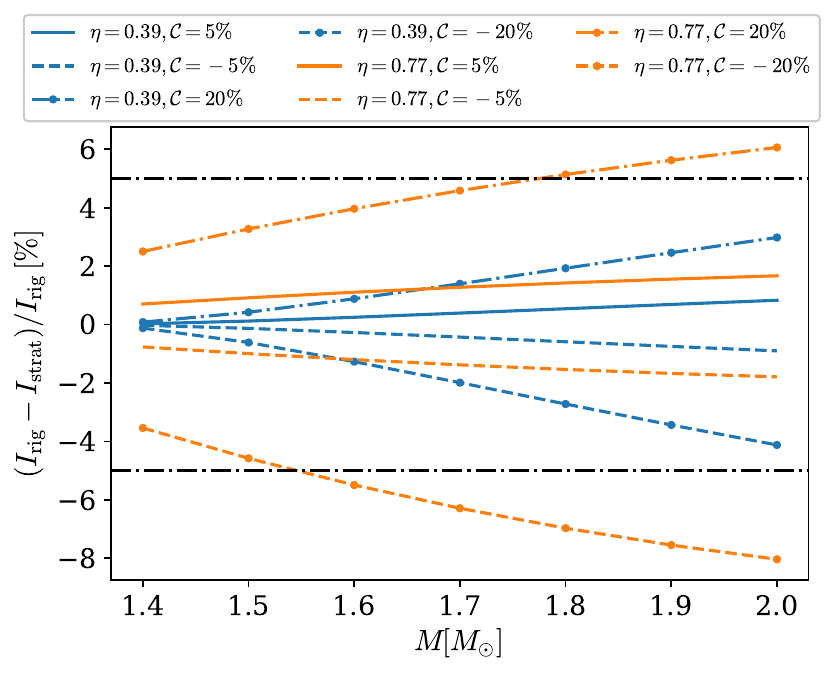}
\caption{$\Delta I/I_{\rm{rig}}$ for hybrid EOS models with $\eta = 0.39$ (weak phase transition) and $\eta=0.77$ (strong phase transition) as a function of the NS's mass assuming an angular velocity jump ${\cal C}$ at the phase transition radius ($R_{\star}$). Several ${\cal C}$ and NS masses have been included. Each color represents a given energy density jump, while each curve dashing relates to a given ${\cal C}\sim [\Omega]^+_-/\Omega^+$. Non-negligible relative differences ($\gtrsim 5\%$) start showing up only for ${\cal C}\sim 20\%$ for these models given that the $R_{\star}$ are at most around $0.8R$ (see Fig. \ref{fig3}).}
\label{fig2}
\end{figure}

\begin{figure}
\centering
\includegraphics[width=\columnwidth]{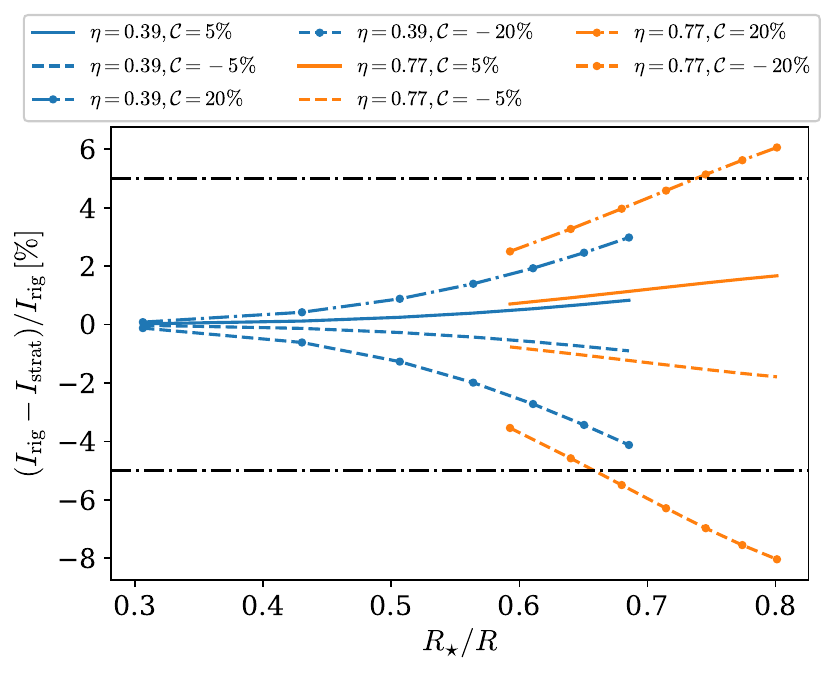}
\caption{$\Delta I/I_{\rm{rig}}$ for hybrid EOS models with $\eta = 0.39$ (weak phase transition) and $\eta=0.77$ (strong phase transition) assuming an angular velocity jump of ${\cal C}$ at the phase transition radius ($R_{\star}$) as a function of $R_{\star}/R$. Several ${\cal C}$ and NS masses have been included, related to the different $R_{\star}/R$ values for a given $\eta$. Each color on the curves relates to a given energy density jump, while each curve shape relates to a given ${\cal C}\sim [\Omega]^+_-/\Omega^+$. Non-negligible relative differences ($\gtrsim 5\%$) start showing up only for $R_{\star}/R\gtrsim 0.8$ and ${\cal C}\sim 20\%$ (because it represents the same models of Fig. \ref{fig2}).}
\label{fig3}
\end{figure}

Figures \ref{fig2} and \ref{fig3} show the results for $\Delta I/I_{\rm{rig}}$ for several choices of ${\cal C}$ and different masses. Qualitatively, the results are similar to the case without phase transitions: relative changes are roughly limited by $[\Omega]^+_-/\Omega^+$, and the deeper the phase transition occurs, the smaller the change in the moment of inertia relative to the rigid case. As is clear from Fig. \ref{fig1}, a relevant aspect is where the jump in rotation occurs in the star. This depends on the star's mass relative to the phase transition mass for a given microphysics of the quark and hadronic phases. Another relevant aspect of $\Delta I/I_{\rm rig}$ as a function of the angular rotation jump depth, as evident in Figs. \ref{fig1} and \ref{fig3}, is its nonlinearity. Instead, in the case of $\Delta I/I_{\rm rig}$ versus the stellar mass, as shown in Fig. \ref{fig2}, an almost linear behavior is observed, mainly because the phase transition does not vary significantly with the mass in the range (1.4-2.0)$M_{\odot}$. The larger the mass, the greater the relative change in the moment of inertia because larger masses have larger transition radii, meaning that the radius where the angular velocity jump occurs is closer to the stellar surface.

%%%%%%%%%%%%%%%%%%%%%%%%%%%%%%%%%%%%%%%%%%%%%%%%%%%%%%
%%%%%%%%%%%%%%%%%%%%%%%%%%%%%%%%%%%%%%%%%%%%%%%%%%%%%%
\section{Discussion and conclusions}
\label{discussion}
%%%%%%%%%%%%%%%%%%%%%%%%%%%%%%%%%%%%%%%%%%%%%%%%%%%%%%
%%%%%%%%%%%%%%%%%%%%%%%%%%%%%%%%%%%%%%%%%%%%%%%%%%%%%%

The internal composition and rotational dynamics of NSs are not directly observable, necessitating theoretical assumptions for predictions. A prevalent assumption is the rigid rotation of NSs, simplifying the model by reducing the number of variables. However, this assumption may introduce biases in astrophysical constraints. Observations of the Sun and other stars suggest that differential rotation is a more natural assumption. However, modeling this requires a specific rotational law, which introduces its own set of challenges. Moreover, as a main-sequence star, the Sun differs significantly from NSs regarding composition, dynamics, and phenomena. Additionally, temperature effects significantly influence the Sun's differential rotation and are less significant in NSs. 

Despite these differences, the Sun is a practical reference point without direct observables for NS internal rotation. Our study shows that non-rigid rotation in NSs introduces systematic uncertainties when calculating their moment of inertia. We adopted a simplified model of non-rigid rotation, characterized by two phases, separated at $r=R_{\star}$, and having different angular velocities. Though rudimentary and predicated on stellar viscosity properties, this model aims to identify angular velocity jumps that significantly impact $I$. Our findings indicate that the maximum relative differences in $I$ correspond to the relative changes in the angular velocity, especially when the discontinuity in rotation occurs near the stellar surface. Considering that future observations could constrain $I$ within $5\%$--$10\%$, our results highlight the potential inaccuracies in assuming rigid rotation for stars.

We now discuss the implications of rotational stratification and changes in $I$ for various stellar observables. Incorporating non-rigid rotation models, which could mitigate many current challenges in understanding stellar dynamics, may help address these issues. Here we focus on the impact of the maximum changes in $I$ on certain observables to assess their systematic errors.

%%%%%%%%%%%%%%%%%%%%%%%%%%%%%%%%%%%%%%%%%%%%%%%%%%%%%
\subsection{Direct measurements of $I$}
%%%%%%%%%%%%%%%%%%%%%%%%%%%%%%%%%%%%%%%%%%%%%%%%%%%%%

From the above, it becomes clear that direct measurement of the NS moment of inertia can be the most relevant probe of the NS interior stratification. Such measurement might soon be fulfilled as pulsar timing precision and data increase and improve \citep{2021PhRvX..11d1050K}, but models must be more accurate. We have just reached the level where higher-order post-Keplerian parameters can be assessed because the double pulsar mass loss cannot be ignored anymore. Among such higher-order parameters, the moment of inertia of a pulsar can be constrained. The results of \citet{2021PhRvX..11d1050K} have constrained the moment of inertia of a pulsar with mass $1.338M_{\odot}$ to $I_{45}\equiv I/(10^{45} \text{g cm}^2)=1.15$--$1.48$ at $95\%$ confidence. It takes into account multimessenger constraints on the radius of NSs. Although the uncertainty about this result is around $10\%$--$20\%$, it is remarkable that such a direct inference can be made. In the future, this uncertainty is expected to decrease significantly.

Another possibility for constraining the moment of inertia of an NS in a binary system is due to measurements of its periastron advance \citep{1988NCimB.101..127D}, which will be possible for some sources \citep{2020ApJ...901..155G}. That is a higher-order spin-orbit effect, and the expected precision for moment-of-inertia measurements is around $10\%$--$20\%$ \citep{2020ApJ...901..155G}. Based on our results, the above accuracies suggest the maximum uncertainties for the rigid rotation of neutron stars. At the same time, it shows that phases rotating with relative differences up to $10\%$--$20\%$ could not be differentiated from those rotating with a uniform angular velocity. However, this stratified rotation could have an impact on other observables, and combined measurements may be able to provide further information. In the best-case scenario, stratified rotation will lead to systematic uncertainties, and they should be duly characterized and not ignored.

Systematic uncertainties to $I$ can have important implications for the constraint of superdense matter in NSs. Suppose there is an intrinsic $\Delta I/I$ due to non-rigid rotation. In that case, it follows that radius inferences will also have uncertainties, and they can be estimated as $\Delta R/R=(1/2)\Delta I/I$ if the mass of an NS is well-constrained. If $\Delta I/I\sim 10\%$, then $\Delta R/R\sim 5\%$. For reference, that already competes with combined multimessenger constraints using ray-tracing techniques for light-curve modeling and tidal deformation constraints \citep{2021ApJ...918L..28M}. We have shown that $\Delta I/I$ around a few percent can already occur for small angular velocity jumps ($\sim 5\%$). Thus, unless the rotation stratification is small, direct measurements of $I$ may constrain superdense matter less strongly than combined multimessenger techniques.

%%%%%%%%%%%%%%%%%%%%%%%%%%%%%%%%%%%%%%%%%%%%%%%%%%%%%
\subsection{Central Compact Objects (CCOs)}
%%%%%%%%%%%%%%%%%%%%%%%%%%%%%%%%%%%%%%%%%%%%%%%%%%%%%

CCOs are a distinctive class of neutron stars, typically found near the centers of young supernova remnants (SNRs)~\citep[see, e.g.,][for a review]{2017JPhCS.932a2006D}. As these remnants evolve from the aftermath of a supernova, they undergo dynamic processes, including the potential accretion of surrounding material.

The assumption of rigid (uniform) rotation, often applied to older neutron stars, may not be valid for these young and dynamically evolving objects. Specifically, it is expected that the accreted layer rotates differently from the rest of the star, at least transiently, over timescales characterized by viscosity in the outermost regions of the NS. Our estimates suggest that this transient period lasts approximately up to $10$ kyr, which is significant for most CCOs given the typical ages of their associated SNRs \citep[see][]{2023IAUS..363...51B}.

Due to the thinness of these accreted layers, which are much smaller than the radius of the NS, our analysis suggests that their moments of inertia might differ considerably from other NSs with similar masses if their layers rotate at different rates compared to the rest of the star. When rotation is considered and is not negligible in ray-tracing techniques used to characterize CCOs (or even rotation-powered pulsars), fluctuations in the moment of inertia may affect the spacetime geometry due to variations in total angular momentum and the quadrupole moment. In addition, fluctuations in $I$ can also influence the inference of the dipolar component of the magnetic field of CCOs. Indeed, $B\propto I^{\frac{1}{2}}$, meaning that $\Delta B/B$=(1/2)$\Delta I/I$ for given values of the star's period and its derivative. For instance, a $\Delta I/I\sim 10\%$ would imply $\Delta B/B\sim 5\%$. Thus, the stratified rotation will not significantly affect the values inferred for the dipolar component of $B$ of CCOs. 

The differential rotation within the accreted layers of CCOs has a broader impact beyond just influencing magnetic fields; it can also affect the emission of X-rays. Specifically, variations in the accretion rate, the angular momentum deposited, and the temperature distribution on the NS surface may lead to measurable fluctuations in X-ray emission. These aspects should be investigated further, as such modulations could provide additional information about the internal structure and dynamical evolution of CCOs. We leave this for future studies.

%%%%%%%%%%%%%%%%%%%%%%%%%%%%%%%%%%%%%%%%%%%%%%%%%
\subsection{Energy budget of Neutron Stars}
%%%%%%%%%%%%%%%%%%%%%%%%%%%%%%%%%%%%%%%%%%%%%%%%%

Here we focus on back-of-the-envelope consequences of the rotational energy of stratified NSs. We have checked numerically that the second term of Eq. \eqref{Erot_strat} is up to around $1\%$ of the first term for all EOSs that we used. Therefore, we will neglect it and consider $E_{\rm{rot}}\simeq (1/2)I(\Omega^+)^2$. A relative change of $10\%-20\%$ in the moment of inertia due to stratified rotation might significantly impact the energy budget of some sources previously classified as non-rotation-powered pulsars. We identified at least five sources that could be explained by rotation alone in the stratified regime by applying the SLy4 equation of state to compute each pulsar's moment of inertia. This was done using the period (\(P\)) and period derivative (\(\dot{P}\)) data from the third catalog of Gamma-Ray Pulsars\footnote{\url{https://fermi.gsfc.nasa.gov/ssc/data/access/lat/3rd_PSR_catalog/}}. For convenience, we define \(\alpha \equiv L_{\gamma}/E_{\text{rot}}\), where \(L_{\gamma}\) is the luminosity and \(E_{\text{rot}}\) is the rotational energy. Thus, a rotation-powered pulsar (RPP) has \(\alpha \leq 1\).

In our analysis, we selected two reference points from the EOS: the fiducial configuration for SLy4, i.e., a mass \(M = 1.45 M_{\odot}\) for which we find a moment of inertia \(I_0 = 1.06 \times 10^{46} \text{ g cm}^2\), and the maximum mass configuration, \(M = 2.04 M_{\odot}\), where we calculated \(I_0 = 6.55 \times 10^{45} \text{ g cm}^2\). Table \ref{table:alpha} displays our results. Considering the fiducial mass, we find that for a $10\%$ increase in the moment of inertia ($I_{+10\%}$), the source J1522--5735 has a value of \(\alpha = 0.99\), indicating it is an RPP, in contrast to \(\alpha = 1.19\) found for \(I_0\). Additionally, when we increase the change in the moment of inertia to $20\%$ ($I_{+20\%}$) for the fiducial mass case, we identify three sources within the rotation-powered zone: J1057--5851, J1522--5735, and J1650--4601. Stratified rotation will not address the issue of RPPs for all pulsars, but it is an aspect that must be considered for energy budget assessments. A reasonable upper limit correction would be $10\%$--$20\%$.

\begin{table}[h]
\centering
\caption{Comparison of \(\alpha\)s for pulsars with rigid-rotation and stratified-rotation moments of inertia}\label{table:alpha}
\begin{tabular}{c|c|c|c}
\multicolumn{4}{c}{$(M,I_0)=(1.45 M_{\odot}, 1.06 \times 10^{46}\, \text{g cm}^2)$} \\ \hline
\textbf{PSR} & \(\alpha(I_0)\) & \(\alpha(I_{+10\%})\) & \(\alpha(I_{+20\%})\) \\ \hline
J1057-5851   & 1.19 & $>1$ & 0.99 \\
J1522-5735   & 1.09 & 0.99 & 0.911 \\
J1650-4601   & 1.11 & $>1$ & 0.92 \\
\multicolumn{4}{c}{} \\ % Spacer
\multicolumn{4}{c}{$(M,I_0)=(2.04 M_{\odot}, 6.55 \times 10^{45}\, \text{g cm}^2)$} \\ \hline
\textbf{PSR} & \(\alpha(I_0)\) & \(\alpha(I_{+10\%})\) & \(\alpha(I_{+20\%})\) \\ \hline
J1429-5911   & 1.03 & 0.94 & 0.86 \\ 
J1817-1742   & 1.04 & 0.95 & 0.87 \\ 
\end{tabular}
\end{table}

%%%%%%%%%%%%%%%%%%%%%%%%%%%%%%%%%%%%%%%%%%%%%%%%

%%%%%%%%%%%%%%%%%%%%%%%%%%%%%%%%%%%%%%%%%%%%%%%%%
\subsection{Mass changes due to stratification}
%%%%%%%%%%%%%%%%%%%%%%%%%%%%%%%%%%%%%%%%%%%%%%%%%

\begin{figure}
\centering
\includegraphics[width=\columnwidth]{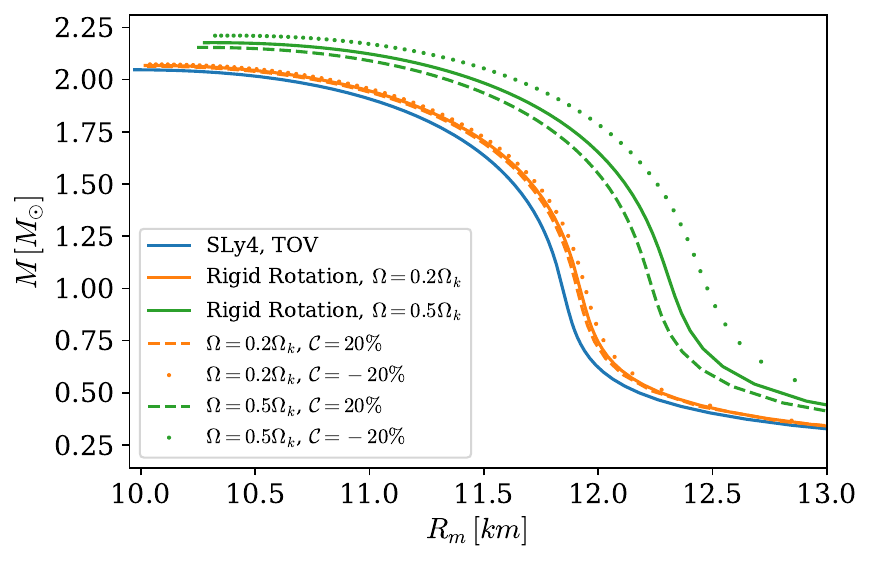}
\caption{Gravitational mass ($M$) versus mean radius ($R_m$) relation, where $R_m = (R_p + 2 R_{\rm eq})/3$ ($R_p$ and $R_{\rm eq}$ are the NSs radius at the pole and equator, respectively). The SLy4 EOS was used as the equation of state of the nonrotating seed. Concerning the rigid rotations, we chose the exemplary cases of $\Omega = 0.2\Omega_k$ and $\Omega = 0.5\Omega_k$, where $\Omega_k=\sqrt{G m(R)/R^3}$ is the Newtonian Keplerian frequency.  For the stars with stratified rotation, we assumed a rotation jump of ${\cal C}=\pm 20\%$ close to the surface $R_*=0.9R$.}
\label{fig5}
\end{figure}

Within the Hartle formalism for slowly rotating stars, the boundary conditions discussed in \ref{boundary} for the first-order quantity $\omega(r)$ will affect all the second-order metric perturbation functions. In particular, the total mass-energy of the rotating star is given by \cite{1968ApJ...153..807H}
\begin{equation}
\begin{split}
     M &= m(R) + m_0(R) + \frac{J^2}{R^3} \equiv  m(R) + \delta M
\end{split}   
\end{equation}
where $m(R)$ is the mass of the non-rotating star and $\delta M$ is the mass increase due to rotation, with a contribution coming from $m_0$, a second-order metric perturbation function, and the rotational energy, which is proportional to $J^2$. The stratification will affect $J$ (in the same way $I$, as shown in \ref{results}) and the function $m_0$ since its differential equation depends on the value of $\omega$ throughout the star. Thus, a stratified NS has a different gravitational mass relative to a rigidly rotating one, as shown in Fig. \ref{fig5} for different $\Omega$s in the reference rigid case ($20\%$ and $50\%$ of the Newtonian Keplerian frequency, $\Omega_k=\sqrt{G m(R)/R^3}$). We chose a situation where the sharp angular velocity change happens close to the star's surface, $R=0.9R$, and a large rotation jump, $|{\cal C}|=20\%$, to obtain the largest differences. For the case, $\Omega=0.2\Omega_k$, stratification contributes little to the mass related to rigid rotation ($\lesssim 1\%$). However, the contribution to the rigid mass due to stratified rotation when $\Omega=0.5\Omega_k$ can be more significant, up to around $10\%$. The above happens essentially because the mass variation $\delta M_{\rm{strat}}$ can differ significantly concerning $\delta M_{\rm{rig}}$ (up to $\approx 45\%$ for the case ${\cal C}=-20\%$ and $R_{\star}$ close to the star surface), but the total mass is still dominated by the non-rotating term $m(R)$, making less relevant the global change in the mass.

%%%%%%%%%%%%%%%%%%%%%%%%%%%%%%%%%%%%%%%%%%%%%%%%%%%%%
\subsection{I-Love-Q universal relations}
%%%%%%%%%%%%%%%%%%%%%%%%%%%%%%%%%%%%%%%%%%%%%%%%%%%%%

Many uncertainties are associated with the NS interior structure, mainly because of our lack of knowledge about the equation of the state of the NS core. Nevertheless, some universal relations between the moment of inertia $I$, the quadrupole moment $Q$, and the tidal Love number $\lambda$ have been discovered some time ago \cite{Yagi:2013bca}. These relations are very powerful from an astrophysical point of view since measuring any of the variables in the trio could lead to direct information about the other two.
The assumptions behind the $I$-Love-$Q$ relations derivation are slow and uniform rotation, small tidal perturbations, and GR gravity. As discussed in this work, letting go of uniform rigid rotation will naturally alter the moment of inertia, as shown previously. But what happens with the quadrupole moment and the universal relations?

The spin-induced quadrupole moment $Q$ is a second-order quantity that will be affected by the boundary condition in $\omega$ through nonlinear terms in the differential equation. The expression for $Q$ is given by \cite{1968ApJ...153..807H}
\begin{equation}
    Q = -\frac{J^2}{m(R)} - \frac{8}{5}A\, m^3(R).
\end{equation}
As we have shown, the first term is sensitive to stratification, hence the second term, since second-order equations are seeded by the solutions of first-order equations in the rotation parameter. 

The universal relations are expressed in terms of the dimensionless moment of Inertia $\bar{I} \equiv I/m(R)^3$ and dimensionless quadrupole moment $\bar{Q} \equiv - Q\,m(R)/J^2$ \cite{Yagi:2013bca}. Since the normalization is done with the nonrotating mass $m(R)$, the relative change due to stratification in the dimensionless moment of inertia will be
\begin{equation}
    \frac{\Delta \bar{I}}{\bar{I}} = \frac{\Delta J}{J} = \frac{\Delta I}{I},
\end{equation}
while the relative change in the dimensionless quadrupole moment is 
\begin{equation}
    \frac{\Delta \bar{Q}}{\bar{Q}} = \frac{\frac{8}{5}\frac{A}{J^2} \, m^4(R) \left(\frac{\Delta A}{{A}} - \frac{2 \Delta J}{J}\right)}{\left(1 + \frac{8}{5}\,A\frac{m^4(R)}{J^2}\right)}.
\end{equation}
One sees that $\Delta \bar{I}$ and $\Delta\bar{Q}$ will not be proportional to each other due to the presence of $\Delta A/A$. Thus, we expect, in general, that the universal relation between $\bar{I}$ and $\bar{Q}$ will no longer hold/will be weakened in the presence of stratified rotation. We leave further examination of that for future work.

%%%%%%%%%%%%%%%%%%%%%%%%%%%%%%%%%%%%%%%%%%%%%%%%%%%%%
\subsection{Gravitational wave production}
%%%%%%%%%%%%%%%%%%%%%%%%%%%%%%%%%%%%%%%%%%%%%%%%%%%%%

For isolated NSs, mountains can be built on their surfaces for a variety of reasons (see \citep{2023ApJ...950..185P} and references therein), and they could lead to the emission of GWs due to a resultant ellipticity (quadrupole moment). It depends on the moment of inertia of the star and is defined as \citep{2005PhRvL..95u1101O,2013PhRvD..88d4004J,2023ApJ...950..185P}
\begin{equation}\label{ellipticity_star}
    \varepsilon \equiv \sqrt{\frac{8\pi}{15}}\frac{Q_{22}}{I},
\end{equation}
where $I$ is the principal moment of inertia, $Q_{22}$ is the $l=m=2$ quadrupole moment of the star. The GW strain is \citep{2019gwa..book.....A}
\begin{equation}
    h=\frac{4\pi \varepsilon I f_{\rm{GW}}^2}{d},
\end{equation}
where $d$ is the source distance and $f_{\rm{GW}}$ is the GW frequency. Thus, from the above equations, $\varepsilon$ (and to a lesser extent $h$ because of the product $\varepsilon I$) will have systematic uncertainties associated with the NS unknown internal rotation. The relative uncertainties of $\varepsilon$ will be proportional to $\Delta I/I$. Uncertainties today are large for $h$, and only upper limits can be set for $\varepsilon$, meaning that $\Delta I/I\lesssim 10\%$--$20\%$ will not be problematic for NS constraints with them. However, in the future, systematics about $I$ will become relevant for GW astronomy. That motivates further studies on the internal rotation of NSs.  

Future missions promise much tighter constraints on GW observables, such as the GW strain, tidal deformations, ellipticities, and quasi-normal modes, which all depend on the moment of inertia. At the same time, when uncertainties for GW observables decrease, it might also be possible to constrain rotation aspects of NSs with them. Statistical studies of GW observables may also constrain where, inside the star, an abrupt rotation change happens. Our analysis suggests that relative changes of a few percent in the moment of inertia happen for $R_{\star}/R\gtrsim 0.6$. If no systematic fluctuation in the observables is found, it would suggest that rotation changes happen much deeper in the star. 

\begin{acknowledgments}
We thank the anonymous referee for their valuable suggestions and comments, which have helped improve this work. J.P.P. is thankful for the financial support of Funda\c c\~ao de Amparo \`{a} Pesquisa e Inova\c c\~ao do Esp\'irito Santo (FAPES) under grant N. 04/2022 and CNPq--Conselho Nacional de Desenvolvimento Cient\'ifico e Tecnol\'ogico--under grant N. 174612/2023-0. J.G.C. is grateful for the support of FAPES (1020/2022, 1081/2022, 976/2022, 332/2023), CNPq (311758/2021-5), and FAPESP (2021/01089-1). T. O. is grateful for the financial support of CAPES and the hospitality of the University of Ferrara. This study was financed in part by the Coordenação de Aperfeiçoamento de Pessoal de Nível Superior - Brasil (CAPES) - Finance Code 001
\end{acknowledgments}

\bibliography{apssamp}% Produces the bibliography via BibTeX.

\end{document}